\documentclass[conference]{IEEEtran}
\usepackage[font=footnotesize]{caption}
\usepackage{caption}
\IEEEoverridecommandlockouts
\usepackage{cite}
\usepackage{amsmath,amssymb,amsfonts}
\usepackage{algorithmic}
\usepackage{algorithm}
\usepackage{graphicx}
\usepackage{textcomp}
\usepackage{xcolor}
\usepackage{stfloats}
\usepackage{multirow}
\usepackage{tabularx}
\usepackage{svg}
\usepackage{subcaption}
\usepackage{diagbox}
\usepackage{mathtools}

\usepackage[skip=2pt]{caption}
\usepackage{subcaption}
\usepackage{etoolbox}
\apptocmd{\thebibliography}{\setlength{\itemsep}{-1.1pt}}{}{}
\makeatletter
\long\def\@makecaption#1#2{%
  \vskip\abovecaptionskip
  \centering
  \small #1: #2\par
  \vskip\belowcaptionskip
}
\makeatother

\def\BibTeX{{\rm B\kern-.05em{\sc i\kern-.025em b}\kern-.08em
    T\kern-.1667em\lower.7ex\hbox{E}\kern-.125emX}}

\begin{document}

\title{Aligning Beam with Imbalanced Multi-modality: A Generative Federated Learning Approach\\

\thanks{This research was supported in part by the Guangdong Provincial Key Laboratory of Future Networks of Intelligence, The Chinese University of Hong Kong, Shenzhen, under Grant No.2022B1212010001-OF08 and in part by National Natural Science Foundation of china under Grant No.62401488, and in part by the Fundamental Research Funds for the Central Universities under Grant 2024ZYGXZR076. (\textit{Corresponding Author: Shijian Gao}).}
}

\author{\IEEEauthorblockN{
    Jiahui Liang\IEEEauthorrefmark{1}\IEEEauthorrefmark{4},
    Miaowen Wen\IEEEauthorrefmark{2},
    Shuoyao Wang \IEEEauthorrefmark{3},
    Yuxuan Liang\IEEEauthorrefmark{1},
    and Shijian Gao\IEEEauthorrefmark{1}\IEEEauthorrefmark{4}
}
\IEEEauthorblockA{\IEEEauthorrefmark{1}The Hong Kong University of Science and Technology (Guangzhou), Guangzhou, China}
\IEEEauthorblockA{\IEEEauthorrefmark{2}South China University of Technology, Guangzhou, China}
\IEEEauthorblockA{\IEEEauthorrefmark{3}College of Electronic and Information Engineering, Shenzhen University, Shenzhen, China}
\IEEEauthorblockA{\IEEEauthorrefmark{3}Guangdong Provincial Key Laboratory of Future Networks of Intelligence, The Chinese University of Hong Kong, \\Shenzhen, China}
}

\maketitle
\begin{abstract}
As vehicle intelligence advances, multi-modal sensing-aided communication emerges as a key enabler for reliable Vehicle-to-Everything (V2X) connectivity through precise environmental characterization. As centralized learning may suffer from data privacy, model heterogeneity and communication overhead issues, federated learning (FL) has been introduced to support V2X. However, the practical deployment of FL faces critical challenges: model performance degradation from label imbalance across vehicles and training instability induced by modality disparities in sensor-equipped agents. To overcome these limitations, we propose a generative FL approach for beam selection (GFL4BS). Our solution features two core innovations: 1) An adaptive zero-shot multi-modal generator coupled with spectral-regularized loss functions to enhance the expressiveness of synthetic data compensating for both label scarcity and missing modalities; 2) A hybrid training paradigm integrating feature fusion with decentralized optimization to ensure training resilience while minimizing communication costs. Experimental evaluations demonstrate significant improvements over baselines achieving 16.2\% higher accuracy than the current state-of-the-art under severe label imbalance conditions while maintaining over 70\% successful rate even when two agents lack both LiDAR and RGB camera inputs. Our code is available at the https://github.com/Jiahui-L/GFL4BS.

\end{abstract}

\section{Introduction}

Accurate beam alignment is crucial for Vehicle-to-Everything (V2X) systems to maintain reliable connections \cite{w1_1, w1_26}. The IEEE 802.11ad standard employs beam sweeping, where the base station (BS) scans predefined directions to find the optimal beam pair without prior channel state information (CSI) knowledge \cite{w1_28}. Under perfect CSI assumptions, channel estimation based methods such as zero-forcing \cite{w1_16} and weighted minimum mean square error \cite{w1_17} perform well. Nevertheless, the continual expansion of antenna arrays and pervasive adoption of latency-sensitive applications significantly exacerbate the estimation overhead in these approaches.

In 6G integrated sensing-communications frameworks, sensor data enables detailed environmental characterization. Hence, sensing-aided methods guide efficient beam alignment. For example, deep learning techniques leveraging RF data have been explored in \cite{w1_21, w1_29, w1_32}. With advancing vehicle intelligence, multi-modal sensors now supplant RF-only solutions due to richer environmental insights \cite{w1_7,w1_30}. While federated learning (FL) addresses privacy concerns inherent to heterogeneous datasets \cite{w1_5,w1_6}, existing FL strategies neglect modality imbalance caused by diverse sensor configurations (e.g., type/performance), leading to incomplete multi-modal inputs and heightened training instability risks. \cite{w1_9} employs a modality‑agnostic approach to accommodate any combination of sensor inputs while \cite{w1_10} adopts vertical FL to capture cross‑modal feature heterogeneity. Nevertheless, both methods fail to address partial data loss or inter‑vehicle label imbalance.

In this study, we recognize a critical issue: in practical V2X systems, the labels and modalities of collected sensor data are imbalanced among vehicles due to varying vehicle trajectories within the BS's coverage area \cite{w1_27} and heterogeneous sensor configurations \cite{w1_4}. Such imbalances can cause the global model to overlook features from underrepresented labels during aggregation and increase the risk of local training failure. To mitigate these imbalances, we propose a generative FL approach for beam selection (GFL4BS).

In GFL4BS, before local training, each vehicle employs an \underline{a}daptive zero-shot \underline{m}ulti-modal \underline{d}ata (AMD) generator to augment imbalanced data associated with specific labels or modalities, ensuring more balanced local updates. Unlike the approach in \cite{w1_25}, which requires vehicles to share data with other vehicles or the BS, the AMD generator creates data solely based on the information from the global model. Additionally, to enhance the expressiveness of synthetic data, a novel spectral-regularized loss function for the AMD generator is carefully designed. This loss function can be rapidly tuned to improve the adaptability of GFL4BS across various imbalanced scenarios. Following \cite{w1_5, w1_6, w1_10}, three prevalent sensors are considered in this work: GPS, LiDAR and RGB camera. Consequently, the neural network model consists of an integration branch and three extractor branches. A distributed training scheme with fused data features is proposed to reduce the risk of local training failures and communication overhead.

Specifically, the BS distributes model branches according to vehicle sensor configurations. Each vehicle updates its model branches using its own sensor data, and then sends them to the BS, which aggregates each branch separately to construct a new global model for the next iteration. Simulations indicate that under label imbalance conditions, GFL4BS enhances Top-1 accuracy compared to benchmarks, closely matching the centralized learning method while maintaining high communication performance. Under modality imbalance conditions, GFL4BS not only achieves high accuracy in model training but also reduces communication overhead during training.

\noindent

\begin{figure*}[!t]
    \captionsetup{belowskip=-2pt}
    \centering
    \begin{subfigure}[b]{0.48\textwidth}
        \centering
        \includegraphics[width=\textwidth]{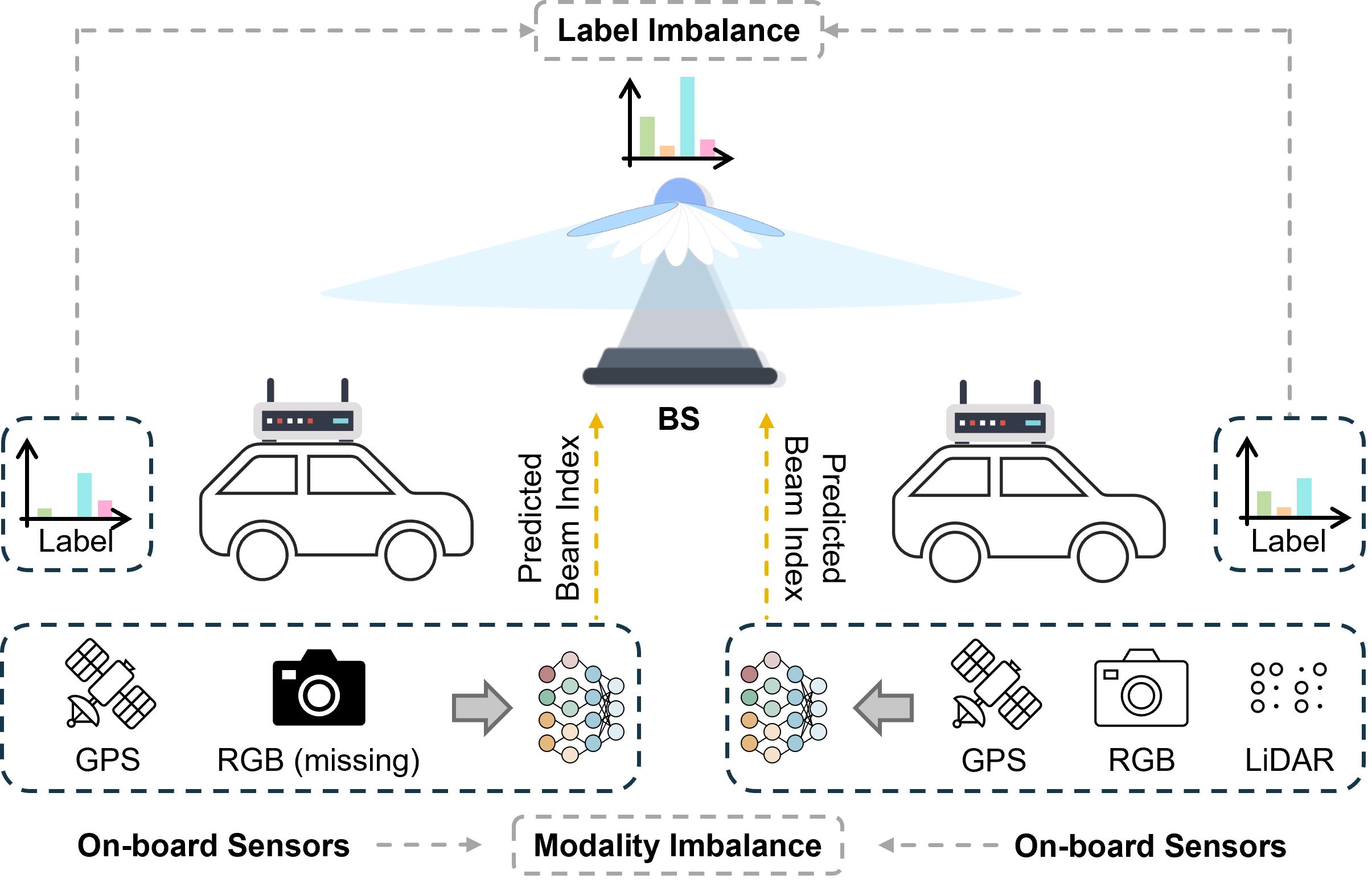}
        \caption{An illustration of system model.}
        \label{fig1a}
    \end{subfigure}
    \hfill
    \begin{subfigure}[b]{0.48\textwidth}
        \centering
        \includegraphics[width=\textwidth]{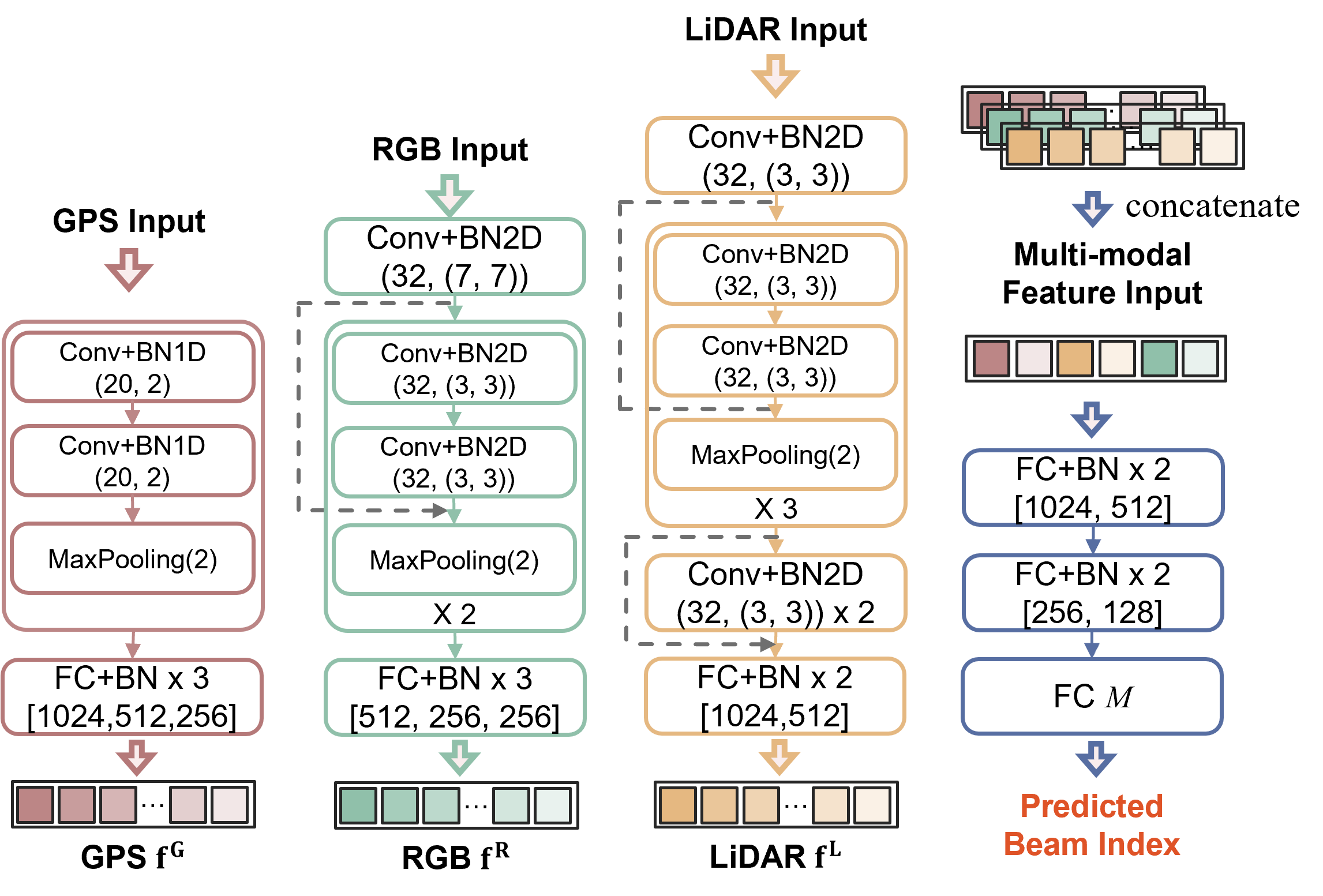}
        \caption{An illustration of modality processing branches.}
        \label{fig1b}
    \end{subfigure}
    \caption{The multi-vehicle system model and modality processing branches.}
    \label{fig:NN_architecture}
\end{figure*}
\vspace{-3pt}
\section{System Model}

As shown in Fig.~\ref{fig1a}, we consider a multi-user multiple-input multiple-output (MIMO) system where the BS, equipped with an $N_t$ uniform array, serves $V$ single-antenna vehicles. A predefined beamforming codebook \(C = \{c^1, c^2, \ldots, c^M\}\) is employed for beam alignment. The BS uses the beamforming vector $\mathbf{w}_{c^m,v} \in \mathbb{C}^{N_t \times 1}$ corresponding to beam $c^m$ to transmit the normalized data symbol $s_v$ to the $v$-th vehicle, subject to the power constraint $\sum_{v=1}^{V}\|\mathbf{w}_{c^m,v}\|^2 \le P$. The received signal at the $v$-th vehicle is expressed as:
{
\setlength{\belowdisplayskip}{3pt}%
\vspace{-3pt}
\begin{equation}
y_v = \mathbf{h}_v^H \mathbf{w}_{c^m,v} s_v + \sum_{i \neq v}^{V} \mathbf{h}_v^H \mathbf{w}_{c^m,i} s_i + n_v,
\label{eq1}
\end{equation}
}where $\mathbf{h}_v \in \mathbb{C}^{N \times 1}$ denotes the channel vector and $n_v \sim \mathcal{CN}(0, \sigma^2)$ represents additive white Gaussian noise. Instead of sweeping beams or exploiting channel estimation, the BS obtains the beamforming vector $\mathbf{w}_{c^m,v}$ by the mapping function \(\mathcal{F}(\cdot)\). The local data of the $v$-th vehicle is represented by $\mathbb{D}_v$. Accordingly, the objective of maximizing the total data rate can be formulated as:
\begin{subequations}\label{eq2}
\begin{align}
\max_{\mathbf{\Theta}} \quad & \sum_{v=1}^{V} \log_2\!\Bigg(1 + \frac{\bigl|\mathbf{h}_v^H \mathbf{w}_{c^m,v}\bigr|^2}{\sum_{i \neq v} \bigl|\mathbf{h}_v^H \mathbf{w}_{c^m,i}\bigr|^2 + \sigma^2}\Bigg) \nonumber \\
\text{s.t.}\quad & \mathbf{w}_{c^m,v} = \mathcal{F}({c^m}) = \mathcal{F}\big({\mathcal{M}}(\mathbb{D}_v;\mathbf{\Theta})\big), \quad \forall v \label{eq2a}\\
& \sum_{v=1}^{V} \|\mathbf{w}_{c^m,v}\|^2 \le P. \label{eq2b}
\end{align}
\end{subequations}
Here, beam index is predicted by \({\mathcal{M}}(\mathbb{D}_v;\mathbf{\Theta})\). The global model \(\mathcal{M}(\cdot)\), parameterized by \(\mathbf{\Theta}\), is trained via FL among the participating vehicles in the set $\mathcal{V}$. In the FL paradigm, local models are collected from the vehicles, aggregated at the BS into a shared global model, and then disseminated back to the vehicles for use in the next training round. 

The modality processing branches shown in Fig.~\ref{fig1b} are specifically tailored to the unique characteristics of each modality. Let \(N_v\) denote the number of samples for each vehicle. Then, the inputs are defined as follows: the GPS data \(\mathbf{x}_v^{\text{GPS}} \in \mathbb{R}^{N_v \times 2}\), the RGB data \(\mathbf{x}_v^{\text{RGB}} \in \mathbb{R}^{N_v \times d_0^{\text{R}} \times d_1^{\text{R}}}\) and the LiDAR data \(\mathbf{x}_v^{\text{L}} \in \mathbb{R}^{N_v \times d_0^{\text{L}} \times d_1^{\text{L}} \times d_2^{\text{L}}}\). Accordingly, the GPS data provides 2D coordinates representing the relative distance between the vehicle and the BS, thereby conveying Line-of-Sight (LoS) information; the RGB data contains rich environmental details, including reflectors and obstacles that contribute to multipath; and the LiDAR data, which is preprocessed into cuboidal regions based on its distance from the BS\footnote{\footnotesize Each cuboidal region is assigned a unique label: blocking obstacles are denoted by \(1\), the transmitter by \(-1\), and the receiver by \(-2\)\cite{w1_6}.}, depicts the spatial structural information of the communication environment. To maintain model stability during both training and inference, every conventional and fully connected layer is followed by a batch normalization (BN) layer. The uni-modal processing branches extract modality features \(\mathbf{f}^{\text{G}}\), \(\mathbf{f}^{\text{R}}\), and \(\mathbf{f}^{\text{L}}\). These features are subsequently concatenated to capture correlations between the raw data and the beam index.

\vspace{-2pt}
\section{Generative Federated Learning approach for Beam Selection}

To address the impact of label and modality imbalances, GFL4BS is proposed. In GFL4BS, vehicles utilize synthetic data generated by the AMD generator to correct model bias in imbalanced scenarios. Additionally, a dedicated distributed training scheme is introduced to mitigate the risk of training failure and reduce transmission overhead.
\vspace{-3pt}
\subsection{Adaptive Zero-shot Multi-modal Data Generator}
To preserve data privacy, it is impractical to train a centralized generator that requires access to data to determine the global distribution in advance. Inspired by \cite{w1_12}, we propose AMD generator that can fulfill multi-modal data generation requirements. During the model training period, each BN layer normalizes the forward activations by computing the mini-batch mean and variance to reduce internal covariate shift and accelerate training, while updating the BN-specific parameters, namely the running mean \(\mu\) and running variance \(\sigma\). These parameters reflect the distribution information of the input data \cite{w1_33}. Therefore, the goal of data generation can be described as finding synthetic data ${\hat{\mathbb{D}}}=\{\hat{\mathrm{d}}^1, \hat{\mathrm{d}}^2, \ldots, \hat{\mathrm{d}}^N\}$ that results in similar statistics to those in the BN layers of the global model. Here, $N$ represents the number of samples and $\hat{\mathrm{d}}^n$ is a tuple consisting $[\hat{x}^{\text{G}, n}, \hat{x}^{\text{R}, n}, \hat{x}^{\text{L}, n}]$. The $\hat{\mu}_l$ and $\hat{\sigma}_l$ of the activation $a_l$ in the forward pass of the $l$-th layer are calculated by:
\begin{figure*}[!t]
    \includegraphics[width=\textwidth]{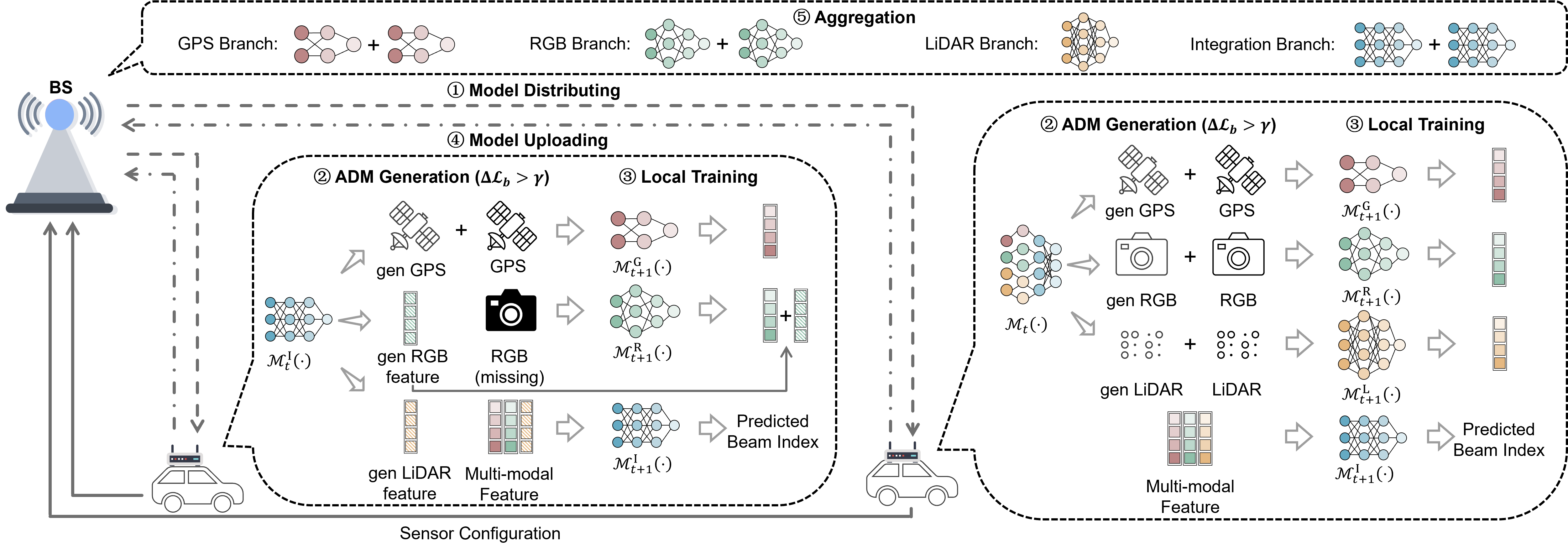}
    \caption{The training flow of the GFL4BS.}
    \label{fig2}
\end{figure*}
{
\setlength{\belowdisplayskip}{5pt}%
\begin{subequations}
\begin{align}
\hat{\mu}_l(\hat{\mathbb{D}}) &= \frac{1}{N} \sum_{\hat{\mathrm{d}}^n \in \hat{\mathbb{D}}} a_l(\hat{\mathrm{d}}^n), \label{eq4a} \\
\hat{\sigma}_l(\hat{\mathbb{D}}) &= \frac{1}{N} \sum_{\hat{\mathrm{d}}^n \in \hat{\mathbb{D}}} \left( a_l(\hat{\mathrm{d}}^n) - \hat{\mu}_l(\hat{\mathbb{D}}) \right)^2. \label{eq4b}
\end{align}
\end{subequations}
}

These data must be labeled for data augmentation. Hence, the cross-entropy function is employed as a label constraint:
\vspace{-3pt}
{
\setlength{\belowdisplayskip}{5pt}%
\begin{equation}
\mathcal{H}(\mathbf{c}, \hat{\mathbf{c}}) = -\frac{1}{N}\sum_{n=1}^{N}\sum_{m=1}^{M} c^{m, n} \log \hat{c}^{m, n},
\label{eq4}
\end{equation}
}where \(M\) is the number of classes, \(c^{m, n}\) is the ground truth for the \(m\)-th class of the \(n\)-th sample, and \(\hat{c}^{m, n}\) is the predicted probability. Each vehicle collects its own necessary labels to construct a hard label set \(\hat{C}_{\text{hard}}\) that constrains the generated data. Notably, if the global model exhibits low accuracy, the quality of the generated data may be compromised. To address this issue, we incorporate $\mathcal{M}_e$ as the evaluation model\footnote{\footnotesize\textsuperscript{}$\mathcal{M}_e$ can be obtained by FedAvg in current imbalanced scenario.} to provide a soft label $\tilde{\mathbf{c}} = \mathcal{M}_e({\hat{\mathbb{D}}})$ to guide the generation process. Then, suppose the global model contains $L^{bn}$ BN layers. The spectral-regularized loss function for the AMD generator is designed as:
\vspace{-4pt}
\begin{equation}
\begin{aligned}
\mathcal{L}_{gen} = \arg\min_{\hat{\mathbb{D}}} \; \sum_{l=1}^{L^{bn}} \big(\|\mu_l - \hat{\mu}_l\|^2 + \|\sigma_l - \hat{\sigma}_l\|^2\big) \\
\quad + \, \mathcal{H}(\hat{\mathbf{c}}_{\text{hard}}, \hat{\mathbf{c}}_{p}) + \,\mathcal{H}(\tilde{\mathbf{c}}, \hat{\mathbf{c}}_{p}).
\label{eq5}
\end{aligned}
\end{equation}
Here, \(\hat{\mathbf{c}}_{p} = \mathcal{M}(\hat{\mathbb{D}})\) represents the prediction of the global model. Gradient descent is employed to solve Eq. \eqref{eq5}. Specifically, the synthetic input is randomly initialized from a Gaussian distribution and iteratively updated to minimize \(\mathcal{L}_{gen}\). Furthermore, Eq. \eqref{eq5} can be adjusted to fulfill the requirement for modality data filling, involving three steps. First, based on the modality type, the initialization is modified from \([\hat{x}^{\text{G},n}, \hat{x}^{\text{R},n}, \hat{x}^{\text{L},n}]\) to a combination of existing raw data and synthetic features \([{x}^{\text{Existed},n}, \hat{f}^{\text{Missing},n}]\). Second, the $L^{bn}$ BN layers in the global model are replaced with $L^{bn}_\text{I}$ BS layers in the integration branch. These two adjustments aim to reduce discrepancies in BN statistics among various extractor branches. Finally, the hard label constraint $\mathcal{H}(\hat{\mathbf{c}}_{\text{hard}}, \hat{\mathbf{c}}_p)$ is replaced with true label constraint $\mathcal{H}(\hat{\mathbf{c}}, \hat{\mathbf{c}}_p)$ because the label $\mathbf{c}$ is known in data filling task.

We observe that after backward gradient propagation, the LiDAR data $\hat{x}_n^{\text{L}}$, which originally comprises four integer values (0, 1, -1, -2), becomes continuous floating-point values. This change results from the update in the $t$-th round:
\begin{equation}
\hat{x}_{t+1}^{\text{L},n} = \hat{x}_{t}^{\text{L},n} - \eta \nabla_{\hat{x}_{t}^{\text{L},n}}\mathcal{L}_{gen}\Big({\hat{x}_{t}^{\text{L},n}}; \{a_{l,t}^{n}\}_{l=1}^{L^{bn}}\Big),
\label{eq6}
\end{equation}
where \(\frac{\partial \mathcal{L}_{gen}}{\partial \hat{x}_{t}^{\text{L},n}}\) is continuous. To address this issue, a binarization mapping function with threshold \(\tau\) is applied, defined as:
\begin{equation}
b(x) =
\begin{cases}
1, & \text{if } x > \tau, \\
0, & \text{if } x \le \tau.
\end{cases}
\label{eq7}
\end{equation}
By doing so, after each update, we embed the receiver and transmitter positions as \(–2\) and \(–1\), respectively, ensuring that the synthetic data format matches the original.

\subsection{Quantification of Label and Modality Imbalances}

Label imbalance arises due to the non-uniform data collection among vehicles, leading to biased local training and suboptimal global model performance. Let \(N_{v}^{m}\) represent the number of samples belonging to label \(m\) on the \(v\)-th vehicle. The local data distribution for the \(v\)-th vehicle can be characterized by \(\mathrm{s}_v = \{ N_v^{1}, N_v^{2}, \ldots, N_v^{M} \}\). The global data distribution can be expressed by aggregating data of all vehicles: \(\mathrm{s}_g = \{ N_g^{1}, N_g^{2}, \ldots, N_g^{M} \}, \quad \text{where} \quad N_g^{m} = \sum_{v=1}^{V} N_v^{m}\). To quantify the label imbalance, we introduce two indicators. 

\textbf{Average Overlap Rate}: The label distribution disparities among vehicles indicate the local label imbalance, which can be expressed as:
\begin{equation}
\zeta = \frac{2}{V(V-1)} \sum_{i=1}^{V-1} \sum_{j=i+1}^{V} \sum_{m=1}^{M} \min\big(\tilde{N}_i^{m}, \tilde{N}_j^{m}\big),
\label{eq8}
\end{equation}
where \(\tilde{N}_i^{m}\) and \(\tilde{N}_j^{m}\) are the normalized ratios of label \(m\) samples at the \(i\)-th and \(j\)-th vehicles, respectively. 

\textbf{Normalized Shannon Entropy}: The difference of \(N^m_g\) reflects the global label imbalance. Define \(\tilde{N}^m_g\) as the globally normalized ratio of label $m$ samples. To illustrate global label imbalance, the Normalized Shannon Entropy is refined as:
\begin{equation}
\epsilon = \frac{-\sum\limits_{m=1}^{M} \tilde{N}_g^{m} \log\big(\tilde{N}_g^{m}\big)}{\log(M)}.
\label{eq9}
\end{equation}

\textbf{Modality Completeness Ratio}: Modality imbalance arises from heterogeneous sensor configurations across vehicles, resulting in risk of local training failure. For vehicle $v$ and modality \(\text{Q} \in \mathcal{Q}\) where $\mathcal{Q} = \{\text{GPS},\, \text{RGB},\, \text{LiDAR}\}$, the collected sensor data is represented as \(\mathbf{x}_v^{\text{Q}} = \{ x_{v}^{\text{Q}} \}^{N_v^{\text{Q}}}\), with the number of samples $N_v^{\text{Q}}$. Denote \(N_v^{\text{Q}^*}\) as the maximum sample count among all modalities for $v$-th vehicle. Then, the modality completeness ratio is described as:
\begin{equation}
\kappa_v^{\text{Q}} = \frac{N_v^{\text{Q}}}{N_v^{\text{Q}^*}}, \quad 0 \le \kappa_v^{\text{Q}} \le 1.
\label{equ10}
\end{equation}
\noindent
Accordingly, each vehicle augments its data with synthetic data generated by AMD generator to drive \(\zeta\), \(\epsilon\) and $\kappa_v^{\text{Q}}$ closer to one. This approach helps mitigate model performance degradation and reduces the risk of training failure by balancing the data distribution across different modalities and vehicles.

\begin{algorithm}[htb]
\caption{Generative Federated Learning Approach}
\label{alg1}
\begin{algorithmic}[1]
\STATE \textbf{Input:} Communication rounds $T$, global model $\mathcal{M}(\cdot)$
\STATE \textbf{Initialization:} Vehicle sends sensor configuration to BS.
\FOR{$t=1, \cdots, T$}
    \STATE Compute global loss decline $\Delta \mathcal{L}_{b}$. \\
    BS sends $\mathbf{\Theta}^{\text{I}}_{v,t}$ and $\mathbf{\Theta}^{\text{Q}}_{v,t},\text{Q} \in \mathcal{Q}_v$ to each vehicle.
    \FOR{each vehicle $v$ \textbf{in parallel}}
        \IF{$\Delta \mathcal{L}_{b} > \gamma$}
            \STATE \textbf{Label Imbalance}:
                \STATE Compute sample shortfall $\Delta \mathbf{N}_v$ by Eq. \eqref{eq15}.
                \STATE Generate synthetic data $\hat{\mathbb{D}}_v$ via Eq. \eqref{eq5}.
            \STATE \textbf{Modality Imbalance}:
                \STATE Generate complete data $\hat{\mathbb{D}}_v$ via modified Eq. \eqref{eq5}.
        \ENDIF
        \STATE Update $\mathbf{\Theta}^{\text{I}}_{v,t}$ and $\mathbf{\Theta}^{\text{Q}}_{v,t},\text{Q} \in \mathcal{Q}_v$ on mixture of $\mathbb{D}_v$ and $\hat{\mathbb{D}}_v$ by Eq. \eqref{eq11a} and Eq. \eqref{eq11a}.
        \STATE Send the updated vehicle parameter $\mathbf{\Theta}^{\text{I}}_{v,t+1}$ and $\mathbf{\Theta}^{\text{Q}}_{v,t+1},\text{Q} \in \mathcal{Q}_v$ to the BS.
    \ENDFOR
    \STATE BS aggregates vehicle parameters to get the updated global model by Eq. \eqref{eq13} and Eq. \eqref{eq14}.
\ENDFOR
\end{algorithmic}
\end{algorithm}

\subsection{Distributed Training Scheme}

In GFL4BS, a distributed training scheme is proposed to reduce the risk of local training failure and communication overhead by delicately fusing raw data and latent features. The whole training process is shown in Fig.~\ref{fig2}. Denote $T$ to be the global communication round. Initially, each vehicle reports its sensor configuration \(\mathcal{Q}_v \subseteq \mathcal{Q}\) to the BS, which then distributes both the corresponding extractor branches \(\{\mathbf{\Theta}^\text{Q}\}_{\text{Q} \in \mathcal{Q}_v}\) and an integration branch \(\mathbf{\Theta}^\text{I}\) to all vehicles. Subsequently, each vehicle updates its integration branch and extractor branches using its local data \(\mathbb{D}_{v}\) according to the following update rules:
\vspace{-10pt}
\setlength{\jot}{1.2em}
\begin{subequations}\label{eq11}
\begin{align}
\text{Integration:}\
\mathbf{\Theta}^\text{I}_{v,t+1}
&= \mathbf{\Theta}^\text{I}_{v,t}
   - \eta\,
     \nabla_{\mathbf{\Theta}^\text{I}_{v,t}}\mathcal{L}_{b}\bigl(\mathbf{\Theta}^\text{I}_{v,t};\,\mathbf f^\text{I}\bigr),
   \label{eq11a}\\
\text{Extractor:}\
\mathbf{\Theta}^\text{Q}_{v,t+1}
&= \mathbf{\Theta}^\text{Q}_{v,t}
   - \eta\,\nabla_{\mathbf{\Theta}^\text{Q}_{v,t}}\mathcal{L}_{b}\bigl(\mathbf{\Theta}^\text{Q}_{v,t};\,\mathbb D_{v}\bigr).
   \label{eq11b}
\end{align}
\end{subequations}
In the above, $\mathbf{f}^{\text{I}}$ represents the integration feature. It is formed by concatenating the extracted features from the available modality data with a zero feature vector \(\mathbf{f}_v^0 \in \mathbb{R}^{L_{\text{Missing}}}\), where \(L_{\text{Missing}}\) denotes the dimension of the missing modality feature. $\mathcal{L}_{b}$ is the loss function used for modal training.

After local training, each vehicle uploads its updated parameters to the BS, which aggregates the parameters separately for each modality. Specifically, the extractor branch parameters for modality $\text{Q} \in \mathcal{Q}$ are aggregated as: 
\vspace{-3pt}
\begin{equation}
\mathbf{\Theta}^{\text{Q}}_{t+1} = \frac{1}{\lvert \mathcal{V}_\text{Q} \rvert} \sum_{v \in \mathcal{V}_\text{Q}} \mathbf{\Theta}^{\text{Q}}_{v,t+1},
\label{eq13}
\end{equation}
where \( \mathcal{V}_\text{Q}\) denotes set of vehicles possessing modality \(\text{Q}\). The integration branch is aggregated across all vehicles:
\begin{equation}
\mathbf{\Theta}^{\text{I}}_{t+1} = \frac{1}{\lvert \mathcal{V} \rvert} \sum_{v \in \mathcal{V}} \mathbf{\Theta}^{\text{I}}_{v,t+1}.
\label{eq14}
\end{equation}
These steps map to lines 13, 14, and 16 in Algorithm 1.

Let $\gamma$ be the threshold for a decline in the loss function that triggers data generation. When $\Delta \mathcal{L}_{\text{b}} > \gamma$, the data generation process is activated. To address label imbalance, each vehicle first examines its local data to identify the label $c^*$ with the highest sample count, denoted by $N_v^*$. Then, the requirement vector $\Delta \mathbf{N}_v \in \mathbb{C}^{M \times 1}$ is computed to quantify the sample shortfall for each label, defined as follows:
\begin{equation}
\bigl[\Delta\mathbf{N}_v\bigr]_m =
\begin{cases}
N_v^* - N_v^{m}, & \text{if } c^m \neq c^*,\\[1mm]
0, & \text{if } c^m = c^*.
\end{cases}
\label{eq15}
\end{equation}

\begin{figure}[t]
\centering
\includegraphics[width=\columnwidth]{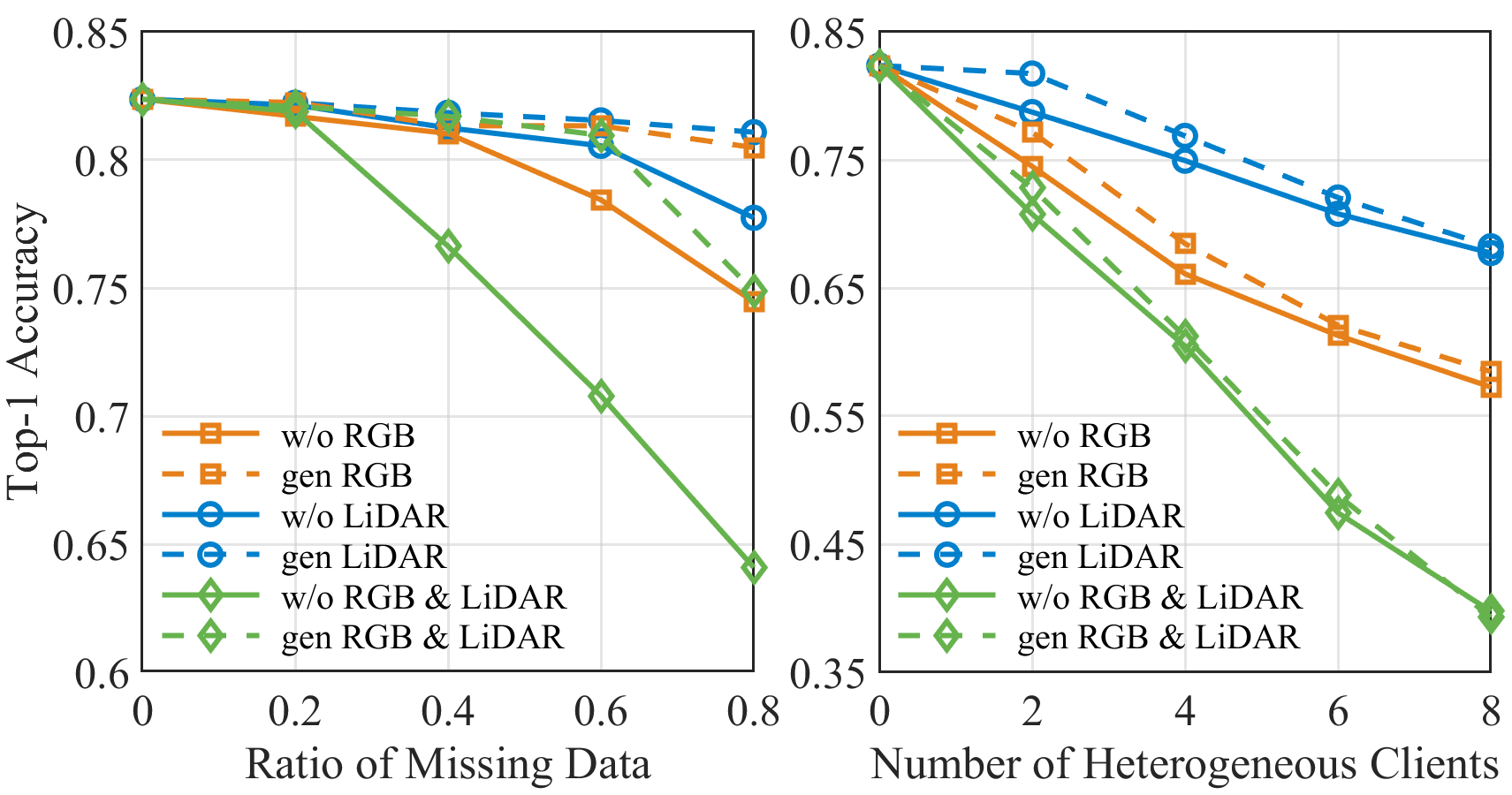}
\caption{Results of GFL4BS in different level modality imbalance.}
\vspace{-3mm}
\label{fig3}
\end{figure}

Subsequently, each vehicle employs an AMD generator to generate data until satisfactory synthetic multi-modal data $\hat{\mathbb{D}}_v$ is obtained. Unlike label imbalance, vehicles facing modality imbalance need only generate one or two specific modal features. Once data generation is complete, the local model is updated using a mixture of $\hat{\mathbb{D}}_v$ and the original local data $\mathbb{D}_v$ until $\Delta \mathcal{L}_{b} > \gamma$ again. Notably, the threshold $\gamma$ is not highly sensitive, since it can be adjusted arbitrarily according to the model’s update step size. Variations in its value do not significantly affect the triggering of the data generation process, thereby ensuring robust performance of the GFL4BS.
\vspace{-5mm}
\begin{table*}[t]
\centering
\caption{Test performance under different levels of imbalance.}
\vspace{1\baselineskip}
\begin{tabularx}{\textwidth}{c *{12}{>{\centering\arraybackslash}X}}
\hline\hline
\multirow{2}{*}{\diagbox{Dataset}{Method}}
  & \multicolumn{3}{c}{GFL4BS}
  & \multicolumn{3}{c}{FedAvg}
  & \multicolumn{3}{c}{FLASH}
  & \multicolumn{3}{c}{CL} \\
\cline{2-13}
  & \textbf{Com} $\uparrow$ & \textbf{Acc} $\uparrow$ & \textbf{Var} $\downarrow$
  & Com & Acc & Var
  & Com & Acc & Var
  & Com & Acc & Var \\
\hline
Original (78.9\%)
  & \textbf{96.8}\% & \textbf{88.5}\% & \textbf{6.4}
  & 94.8\%         & 82.4\%         & 7.5
  & 91.9\%         & 72.3\%         & 12.7
  & 96.9\%         & 89.0\%         & 4.5 \\
\underline{\textbf{L}} (44.3\%)
  & \textbf{95.9}\% & \textbf{80.7}\% & \textbf{46.1}
  & 93.8\%         & 76.3\%         & 82.5
  & 89.3\%         & 66.1\%         & 72.6
  & 96.9\%         & 89.0\%         & 3.9 \\
\underline{\textbf{M}} (35.1\%)
  & \textbf{94.8}\% & \textbf{76.7}\% & \textbf{78.4}
  & 91.7\%         & 72.1\%         & 125.3
  & 89.1\%         & 63.0\%         & 118.3
  & 96.9\%         & 89.0\%         & 3.7 \\
\underline{\textbf{H}} (24.7\%)
  & \textbf{93.3}\% & \textbf{74.3}\% & \textbf{99.2}
  & 90.4\%         & 70.0\%         & 249.1
  & 88.0\%         & 56.9\%         & 163.0
  & 96.9\%         & 89.0\%         & 3.3 \\
\hline
\end{tabularx}
\label{tab1}
\end{table*}


\section{Simulations}

\subsection{Experiment Setting}\label{4A}


Experiments are conducted using the FLASH dataset \cite{w1_5}. This dataset was collected from a vehicle equipped with multi-modal sensors and IEEE 802.11ad Talon Routers operating in the 60 GHz band. It employs a default codebook consisting of 34 sectors. It encompasses both LoS and Non‑Line‑of‑Sight (NLoS) scenarios and includes obstacles, such as pedestrians and vehicles. Data collection is performed across 10 trials and can be interpreted as accommodating different vehicles.

1) \textbf{Label-Imbalanced Dataset:} 
The original dataset exhibits the label imbalance, with average overlap rate: $\zeta=\textbf{78.9\%}$. To further evaluate GFL, there are three datasets with varying degrees of imbalance, namely Low (\underline{\textbf{L}}), Medium (\underline{\textbf{M}}), and High (\underline{\textbf{H}}). The labeled data are first ranked by volume and then split into two equal groups (top 50\% and bottom 50\%). One of these groups is randomly selected for data removal based on a 7:3 probability ratio. Specifically, the numbers of data points removed for the \underline{\textbf{L}}, \underline{\textbf{M}}, and \underline{\textbf{H}} datasets are 6, 9 and 12, respectively. These labeled data are transferred to the next client until the final client receives the additional data. This strategy not only increases the imbalance level but also mitigates any negative effects from reducing dataset size. To minimize the influence of random label selection, the set of removed labels in the higher-imbalance datasets is constructed based on those of the lower-imbalance dataset. The $\zeta$ values for the datasets are \textbf{44.3\%}, \textbf{35.1\%}, and \textbf{24.7\%}.

\begin{figure}[t]
\centering
\includegraphics[width=\columnwidth]{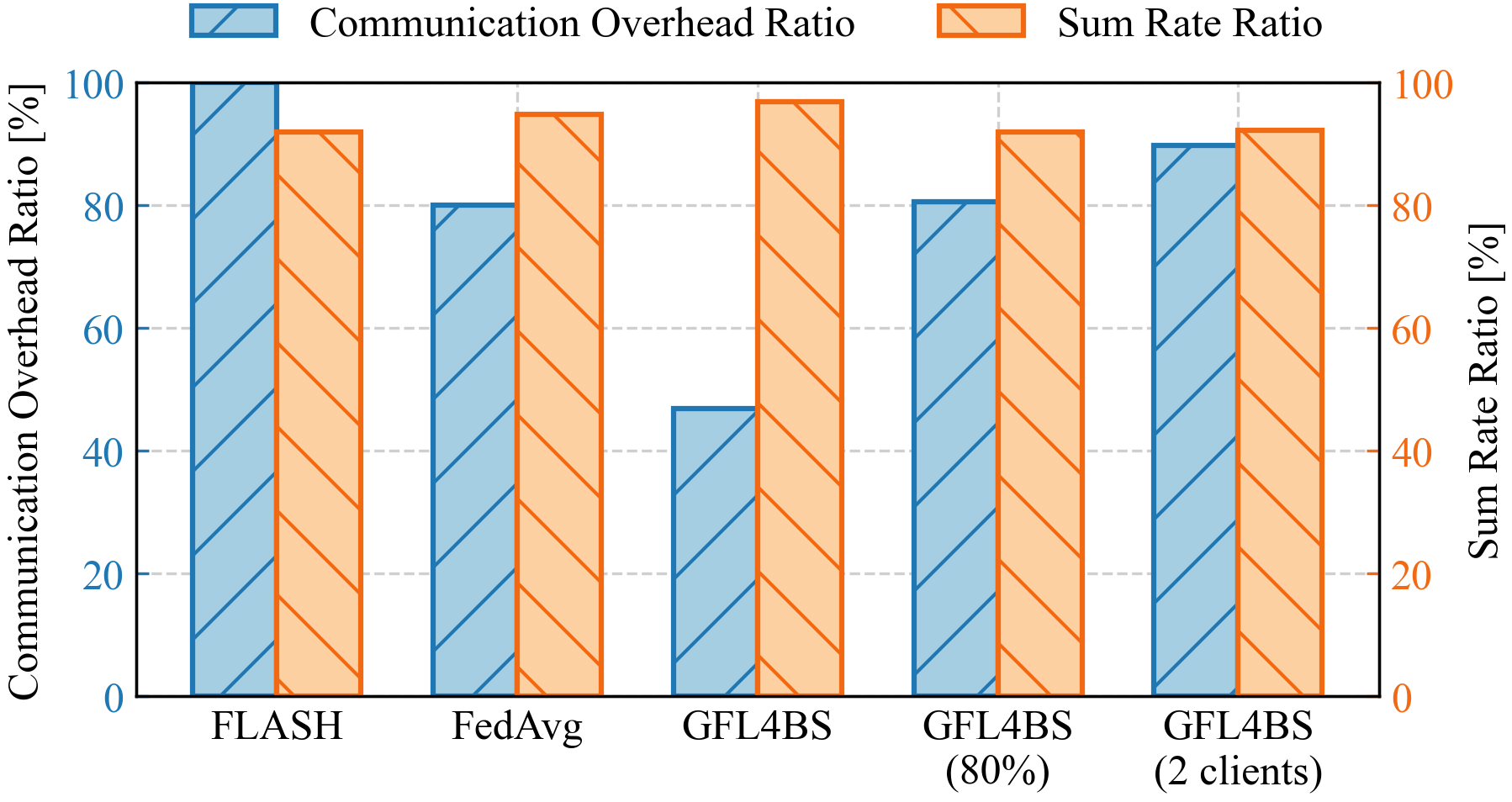}
\caption{Comparison of sum rate and communication overhead ratios for FedAvg, FLASH, and GFL4BS.}
\label{fig4}
\end{figure}

2) \textbf{Modality-Imbalanced Dataset:} To mimic modality imbalance, we employ different masking strategies that simulate varying degrees of missing data. For partial missing data, each vehicle randomly loses 20\%, 40\%, 60\%, or 80\% of its modality data. To simulate complete missing data, we randomly omit the entire modality data from 2, 4, 6, or 8 vehicles. Given that GPS is a standard sensor in vehicles and is expected to be consistently available, these strategies are specifically applied only to the RGB and LiDAR modalities. 

3) \textbf{Implementation Details:} Table ~\ref{tab2} illustrates the hyperparameter configuration for training and data generation. Each vehicle's local dataset is partitioned into three subsets: 80\% for local training, 10\% for local validation, and the remaining 10\% aggregated across all vehicles for the global test.

\begin{table}[t]
\centering
\caption{Hyperparameter Configuration.}
\vspace{1\baselineskip}
\begin{tabularx}{\columnwidth}{>{\centering\arraybackslash}X >{\centering\arraybackslash}X}
\hline\hline
\textbf{Parameter} & \textbf{Value} \\
\hline
Optimizer & Adam \\
Batch Size & 128 \\
Learning Rate & \(10^{-4}\) \\
Epochs at Generation & 500 \\
Epochs at Local Training & 5 \\
Epochs at Global Training & 500 \\
\hline
\end{tabularx}
\label{tab2}
\vspace{-5mm}
\end{table}
\vspace{-0.5mm}
\subsection{Benchmarks}
We compare three representative benchmarks as follows:
\begin{itemize}
    \item \textbf{Centralized Learning (CL):} All vehicles transmit their local training data to the BS, which then trains a global model on the aggregated dataset. This approach requires a control node at the BS to handle the high data volume.

    \item \textbf{Federated Averaging (FedAvg) \cite{w1_24}:} Each vehicle updates local model only on its own data. During global aggregation, local model are sent to the BS, which then distributes the aggregated model back to all vehicles.

    \item \textbf{Federated Learning for Automated Selection of High-band mmWave Sectors (FLASH) \cite{w1_5}:} BS collects updated local models from each vehicle, aggregating them into a global model. An orchestrator at the BS selects one model branch and returns it to the vehicles.

\end{itemize}
\subsection{Performance Comparison with Label Imbalance}


Table ~\ref{tab1} presents the communication performance, Top‑1 accuracy on the global test, and the variance of local test performance for GFL4BS and the benchmarks. The communication performance is measured by the sum rate ratio relative to beam sweeping, as defined in Eq.~\eqref{eq2}. The Top-1 accuracy demonstrates the model’s performance on new vehicles. The variance among vehicles serves as a key indicator of the model's balanced performance \cite{w1_12}. A lower variance signifies that the model offers a more balanced solution. The benefits of the GFL4BS become more pronounced with increasing label imbalance. On the original FLASH dataset, GFL4BS achieves a 16.2\% Top-1 accuracy improvement compared to FLASH and exhibits communication performance comparable to CL.


\subsection{Performance Comparison with Modality Imbalance}

Fig.~\ref{fig3} presents the performance of GFL4BS under varying levels of modality imbalance. The Top-1 accuracy of the evaluation model serves as a baseline metric. The global model’s performance declines as more modality data becomes partially missing. When each vehicle loses 80\% of its RGB\&LiDAR data, the Top-1 accuracy falls to 64.1\%. GFL4BS boosts the Top-1 accuracy to 74.9\% by effectively reconstructing missing features. At lower levels of data loss, GFL4BS performs comparably to lossless models. However, in scenarios where entire modalities are missing, GFL4BS shows little improvement due to limited inter-modality information. Additionally, it fails to yield any gains when data is missing from eight vehicles.

\subsection{Communication Performance}

Fig.~\ref{fig4} illustrates communication performance results. GFL4BS yields a 96.8\% sum rate ratio at an average overlap rate of \(\zeta\)=78.9\% and a 92.9\% sum rate ratio when 80\% of the RGB and LiDAR data are missing. Communication overhead during training is quantified by defining an overhead ratio relative to communication performance of FLASH. As shown, under label imbalance, GFL4BS requires fewer resources because additional data reduces the number of communication rounds. FLASH does not exhibit an advantage, since fewer local training epochs result in slower convergence. The two vehicles lacking LiDAR and RGB cameras transmit the GPS extractor and integration branches to the BS, which account for approximately 25\% of the model. Consequently, GFL4BS reduces communication overhead per training round by 15\%.

\section{Conclusions}

In this paper, we proposed a generative federated learning approach for beam selection. Our approach addressed challenges of label and modality imbalances in multi-modal FL by integrating an adaptive zero-shot multi-modal data generator, which effectively rectifies the abnormal data distributions. Additionally, we introduced a distributed training scheme that fuses data and features to reduce the risk of local training failures caused by modality imbalance.  Simulations validate that GFL4BS not only improves Top-1 accuracy compared to the benchmarks but also efficiently alleviates the adverse effects of varying levels of label and modality imbalances.

\bibliographystyle{IEEEtran}
\bibliography{reference}

\end{document}